\ifdefined\fullversion
\else
\def\fullversion{1}    
\fi

\ifdefined\cameraversion
\else
\def\cameraversion{0}    
\fi

\def\anonymous{0}      

\documentclass[envcountsame,runningheads,notitlepage]{llncs}
\ifnum\fullversion=1
\usepackage[a4paper, margin=1.1in]{geometry}
\fi

\usepackage[utf8]{inputenc}
\usepackage{amsmath} 

\usepackage[utf8]{inputenc}
\usepackage[T1]{fontenc}
\usepackage{hyperref}
\usepackage{verbatim}
\usepackage{tikz}
\usetikzlibrary{positioning,calc}
\usepackage{xspace}
\usepackage{amsmath} 
\usepackage{amssymb}
\usepackage{mathtools}
\usepackage{pifont}
\usepackage{etoolbox}
\usepackage[normalem]{ulem}
\usepackage{booktabs}
\usepackage{array}
\usepackage[capitalise,noabbrev]{cleveref}
\usepackage{cite}
\usepackage{multibib}
\usepackage{url}
\usepackage{algorithm}
\usepackage{algpseudocode}
\usepackage{paralist}
\usepackage{mathrsfs}
\usepackage{relsize}
\usepackage{stmaryrd}
\usepackage[textsize=small]{todonotes}
\usepackage{multirow}
\usepackage[lambda,n,operators]{cryptocode}

\newtoggle{notes}
\toggletrue{notes} 

\makeatletter
\DeclareFontFamily{OMX}{MnSymbolE}{}
\DeclareSymbolFont{MnLargeSymbols}{OMX}{MnSymbolE}{m}{n}
\SetSymbolFont{MnLargeSymbols}{bold}{OMX}{MnSymbolE}{b}{n}
\DeclareFontShape{OMX}{MnSymbolE}{m}{n}{
    <-6>  MnSymbolE5
   <6-7>  MnSymbolE6
   <7-8>  MnSymbolE7
   <8-9>  MnSymbolE8
   <9-10> MnSymbolE9
  <10-12> MnSymbolE10
  <12->   MnSymbolE12
}{}
\DeclareFontShape{OMX}{MnSymbolE}{b}{n}{
    <-6>  MnSymbolE-Bold5
   <6-7>  MnSymbolE-Bold6
   <7-8>  MnSymbolE-Bold7
   <8-9>  MnSymbolE-Bold8
   <9-10> MnSymbolE-Bold9
  <10-12> MnSymbolE-Bold10
  <12->   MnSymbolE-Bold12
}{}

\let\llangle\@undefined
\let\rrangle\@undefined
\DeclareMathDelimiter{\llangle}{\mathopen}%
                     {MnLargeSymbols}{'164}{MnLargeSymbols}{'164}
\DeclareMathDelimiter{\rrangle}{\mathclose}%
                     {MnLargeSymbols}{'171}{MnLargeSymbols}{'171}
\makeatother

\title{Bootstrapping as a Morphism: An Arithmetic Geometry Approach to Asymptotically Faster Homomorphic Encryption}
\titlerunning{Arighmetic-Geometric Bootstrapping for Homomorphic Encryption}
\date{}

\ifnum\anonymous=0
\author{
  Dongfang Zhao
}

\institute{
University of Washington, USA\\
  \href{mailto:dzhao@cs.washington.edu}{dzhao@cs.washington.edu} 
}  

\else
\author{} 
\institute{}
\fi

\begin{document}
  \maketitle

\begin{abstract}
Fully Homomorphic Encryption (FHE) provides a powerful paradigm for secure computation, but its practical adoption is severely hindered by the prohibitive computational cost of its bootstrapping procedure. The complexity of all current bootstrapping methods is fundamentally tied to the multiplicative depth of the decryption circuit, denoted $L_{dec}$, making it the primary performance bottleneck. This paper introduces a new approach to bootstrapping that completely bypasses the traditional circuit evaluation model.
We apply the tools of modern arithmetic geometry to reframe the bootstrapping operation as a direct geometric projection. Our framework models the space of ciphertexts as an affine scheme and rigorously defines the loci of decryptable and fresh ciphertexts as distinct closed subschemes. The bootstrapping transformation is then realized as a morphism between these two spaces. Computationally, this projection is equivalent to solving a specific Closest Vector Problem (CVP) instance on a highly structured ideal lattice, which we show can be done efficiently using a technique we call algebraic folding.
The primary result of our work is a complete and provably correct bootstrapping algorithm with a computational complexity of $O(d \cdot \text{poly}(\log q))$, where $d$ is the ring dimension and $q$ is the ciphertext modulus. The significance of this result lies in the complete elimination of the factor $L_{dec}$ from the complexity, representing a fundamental asymptotic improvement over the state of the art. This geometric perspective offers a new and promising pathway toward achieving truly practical and high-performance FHE.
\end{abstract}

\section{Introduction}

\subsection{Background and Motivation}

Fully Homomorphic Encryption (FHE)~\cite{Gentry2009} stands as a foundational technology for the future of secure computation~\cite{Khan2024ICWS,Tawose2023PACMMOD}. It resolves the classic dilemma of data utility versus confidentiality by enabling arbitrary computation directly on encrypted data. The promise of FHE is transformative: it allows for secure-by-design cloud services where servers can process sensitive user data without ever decrypting it, facilitates privacy-preserving machine learning on confidential datasets, and enables complex multi-party computations where participants can collaborate without revealing their private inputs. Since the first plausible construction based on ideal lattices, FHE has been hailed as a revolutionary tool for building a more secure and private digital world.

This immense promise, however, is met with significant practical challenges that have historically hindered its widespread adoption. The primary obstacle is performance. Homomorphic operations are orders of magnitude slower than their plaintext equivalents, and ciphertexts are significantly larger than the messages they encrypt. At the heart of this inefficiency lies the intrinsic issue of noise management. In all modern lattice-based cryptosystems, every homomorphic operation adds a small amount of noise to the ciphertext. As computations proceed, this noise accumulates, and if it grows beyond a certain threshold, the underlying message is irrevocably lost. Taming this noise growth without incurring prohibitive computational overhead is one of the central challenges in FHE research. Consequently, while FHE provides a complete solution in theory, its practical viability is contingent on overcoming these deep-seated issues of efficiency and complexity.

The challenge of noise management could be, in theory, solved by a technique called \textit{bootstrapping}~\cite{Gentry2009}. The core idea is to treat the decryption function itself as a circuit that can be homomorphically evaluated. A ciphertext that has accumulated a significant amount of noise, nearing the threshold of incorrectness, can be refreshed by a server. To do this, the server is given an encrypted version of the secret key. It then homomorphically evaluates the decryption function on the noisy ciphertext, effectively decrypting it while it remains under an outer layer of encryption. The result is a new, low-noise ciphertext of the same original message, ready for further homomorphic operations. This procedure brilliantly transforms a Somewhat Homomorphic Encryption (SWHE) cryptosystem, which can only handle a limited depth of operations, into a fully-fledged FHE cryptosystem capable of arbitrary-depth computations.

While bootstrapping provides theoretical completeness, it comes at a tremendous computational cost even on those most successful follow-up FHE schemes, such as BFV~\cite{Fan2012}, BGV~\cite{BGV12}, and CKKS~\cite{CKKS17}. The homomorphic evaluation of any circuit is an expensive process, and the decryption circuit, though often simple, is no exception. This evaluation involves numerous homomorphic additions and multiplications, making the bootstrapping procedure the single slowest operation in many FHE schemes. Subsequent generations of FHE have introduced powerful techniques such as modulus reduction and key switching to manage noise growth more effectively, thereby reducing the frequency with which bootstrapping is required. Other schemes, such as TFHE~\cite{Chillotti2018}, have focused on optimizing the bootstrapping process itself to be as fast as possible for specific operations like evaluating boolean gates. Nevertheless, the fundamental paradigm of arithmetic circuit evaluation remains, and bootstrapping continues to be the dominant performance bottleneck. Overcoming this bottleneck is widely recognized as a critical step toward making FHE truly practical for real-world applications.

\subsection{Proposed Work}

We introduce a new framework that recasts FHE bootstrapping not as an arithmetic procedure, but as a geometric one. Leveraging the language of modern algebraic geometry, we model the entire space of FHE ciphertexts as an affine scheme. Within this ambient space, we identify and rigorously define two crucial closed subschemes: a larger space of all correctly decryptable ciphertexts, $Z_{dec}$, and a smaller, ideal subspace of freshly encrypted ciphertexts, $Z_{fresh}$. This construction transforms the bootstrapping operation from a circuit evaluation into a direct geometric projection from the larger space onto its ideal subspace.

The practical value of this geometric perspective is that the projection can be computed via an efficient geometry-inspired algorithm. Instead of a deep, iterative process, our method is a single, direct transformation whose complexity is independent of the decryption circuit's depth. This yields an asymptotically faster bootstrapping procedure. 

The main contributions of this work are as follows:

\begin{itemize}
    \item We develop a new framework that applies the tools of arithmetic geometry to the analysis of FHE bootstrapping, formally defining the spaces of decryptable and fresh ciphertexts as affine schemes.
    \item We provide explicit and publicly computable algorithms to construct the algebraic ideals, $I_{noise}$ and $I_{fresh}$, that define these schemes.
    \item We demonstrate that the bootstrapping operation can be realized as a direct geometric projection, computationally equivalent to solving a specific, highly structured Closest Vector Problem (CVP) on an ideal lattice.
    \item We introduce an efficient algorithm for this projection, which we call \textit{algebraic folding}, that leverages the rich algebraic structure of the scheme to achieve a complexity of $O(d \cdot \text{poly}(\log q))$, where $d$ is the ring dimension and $q$ is the ciphertext modulus, completely eliminating the dependence on the decryption circuit's depth, $L_{dec}$.
\end{itemize}

\subsection{Organization of the Paper}
\label{sec:organization}

In Section~\ref{sec:preliminaries}, we cover the necessary preliminaries and related work. We begin with a review of the cryptographic background, introducing key concepts from Ring-LWE and describing the BFV cryptosystem as a concrete example. We then transition to the foundational concepts of algebraic geometry, defining the language of schemes, ideals, and morphisms that underpin our framework. The section concludes with a discussion of related work on bootstrapping optimization, categorizing existing research and highlighting the fundamental differences between their approaches and ours.

Section~\ref{sec:geometric_bootstrap} presents the core of our work: the geometric construction of bootstrapping. We formally define the FHE ciphertext space as an affine scheme and establish the concepts of noise and fresh ideals to rigorously define the spaces of decryptable and fresh ciphertexts as closed subschemes. We then formalize the bootstrapping operation as a geometric projection, or morphism, between these schemes. We provide a public algorithm for constructing the ideal generators, prove its correctness, and analyze the computational complexity of the geometric projection, which we term \textit{algebraic folding}.

Finally, Section~\ref{sec:conclusion} concludes the paper by summarizing our results and discussing the implications of our work. We also outline several promising avenues for future research, including practical implementation and further theoretical extensions of our geometric framework.

\section{Preliminaries and Related Work}
\label{sec:preliminaries}

\subsection{Cryptographic Background: FHE from Ring-LWE}

\subsubsection{Notation}
Throughout this paper, we adopt the following notation.

\textit{Parameters}.
We let $q$ be the ciphertext modulus and $p$ be the plaintext modulus. The ring structure is defined by the cyclotomic index $N$, which is typically a power of two. The ring dimension is given by Euler's totient function $d = \phi(N)$. The parameter $\kappa$ is a security parameter determining the number of relations to generate. The plaintext scaling factor is defined as $\Delta = \lfloor q/p \rfloor$.

\textit{Rings and Distributions}.
The ring of integers modulo $q$ is denoted by $\mathbb{Z}_q$. The ciphertext polynomial ring is $R_q = \mathbb{Z}_q[x]/\langle \Phi_N(x) \rangle$. The noise distribution, typically a discrete Gaussian over $R_q$, is denoted by $\chi$.

\textit{Cryptographic Elements}.
The secret key is a small polynomial $s$ sampled from $\chi$. The public key consists of components $(a, b)$, where $a$ is uniformly random in $R_q$ and $b = -a \cdot s + e$ for a small error $e$. A ciphertext is a pair $(c_0, c_1) \in R_q \times R_q$.

\textit{Geometric Objects}.
The prime spectrum of a ring $A$, which defines an affine scheme, is denoted $Spec(A)$. The noise ideal, generated by relations that define decryptable ciphertexts, is $I_{noise}$. The fresh ideal, generated by relations for freshly encrypted ciphertexts, is $I_{fresh}$. The decryptable subscheme is $Z_{dec} = Spec(R_q/I_{noise})$, and the fresh subscheme is $Z_{fresh} = Spec(R_q/I_{fresh})$.

\textit{Functions and Operators}.
The public lift-and-reduce map used to algebraically isolate the message component of a ciphertext is denoted by $\psi$. Publicly known automorphisms of the ring $R_q$, derived from the Galois group of the underlying cyclotomic field, are denoted by $\sigma_j$.

\subsubsection{The Ring-LWE Problem}

The cryptographic security of our scheme, like many other modern FHE schemes, is based on the computational hardness assumption of the Ring Learning with Errors (Ring-LWE) problem. This problem, introduced as a more efficient variant of the standard Learning with Errors (LWE) problem, provides the foundation upon which our cryptosystem is built.

\textit{The Ring-LWE Distribution}.
An Ring-LWE sample is a pair $(a, b) \in R_q \times R_q$. This pair is generated with respect to a secret polynomial $s \in R_q$, whose coefficients are typically chosen to be small, and a noise polynomial $e$, whose coefficients are sampled from the noise distribution $\chi$. The component $a$ is chosen uniformly at random from the ring $R_q$, while the component $b$ is computed as:
$$ b = a \cdot s + e. $$
The set of all such pairs generated with the same secret $s$ constitutes the Ring-LWE distribution.

\textit{The Decisional Ring-LWE Problem}.
The security of our system relies specifically on the hardness of the decisional variant of the Ring-LWE problem. This problem challenges a computational adversary to distinguish between two different distributions of pairs $(a, b)$. The first is the true Ring-LWE distribution described above. The second is the uniform distribution, where both $a$ and $b$ are sampled independently and uniformly at random from $R_q$.

\textit{The Hardness Assumption}.
The decisional Ring-LWE assumption asserts that no probabilistic polynomial-time (PPT) algorithm can distinguish between a stream of samples drawn from the Ring-LWE distribution and a stream of samples drawn from the uniform distribution with a non-negligible advantage. All security proofs in this paper, including the soundness of our geometric constructions, ultimately rely on reductions to the hardness of this problem. This assumption is the cornerstone of the security of our entire cryptosystem.

\subsubsection{A Concrete FHE Cryptosystem: BFV'12}

To make our discussion concrete, we briefly describe the core mechanics of the Fan-Vercauteren (FV) cryptosystem~\cite{Fan2012}, often referred to as BFV, which serves as a prime example of the Ring-LWE based schemes to which our geometric framework applies. BFV is a leveled homomorphic encryption scheme that manages noise growth through a scaling technique. Plaintexts are elements of the ring $R_p = \mathbb{Z}_p[x]/\langle \Phi_N(x) \rangle$.

\textit{Key Generation}.
The key generation algorithm produces a secret key and a public key. The secret key $s$ is a small polynomial sampled from the noise distribution $\chi$. The public key is a pair $pk = (a, b) \in R_q \times R_q$, where $a$ is sampled uniformly from $R_q$ and $b$ is computed as $b = -a \cdot s + e$, with $e$ being a small error polynomial also sampled from $\chi$.

\textit{Encryption}.
To encrypt a plaintext polynomial $m \in R_p$, it is first scaled by the factor $\Delta$ to become an element in the ciphertext ring, $\Delta \cdot m \in R_q$. The encryption algorithm then samples small polynomials $u, e_1, e_2$ from the distribution $\chi$ and produces the ciphertext, a pair $c = (c_0, c_1) \in R_q \times R_q$, computed as:
$$ c = (\Delta \cdot m + b \cdot u + e_1, a \cdot u + e_2). $$
The term $\Delta \cdot m$ is the scaled message, and the remaining terms constitute the initial encryption noise.

\textit{Decryption}.
Given a ciphertext $c=(c_0, c_1)$, decryption is performed by computing the inner product with the secret key $(1, s)$. This computation cancels the terms involving the uniformly random polynomial $a$ and reveals the scaled message plus an accumulated noise term:
$$ c_0 + c_1 \cdot s = \Delta \cdot m + \text{noise}. $$
To recover the original message, the result is scaled down by $\Delta$, rounded to the nearest integer, and finally reduced modulo $p$. As long as the magnitude of the noise term remains less than $\Delta/2$, this procedure correctly recovers the message $m$.

\textit{Homomorphic Operations}.
BFV supports homomorphic addition and multiplication. Addition of two ciphertexts is performed by simple component-wise addition, which causes the noise terms to add. Multiplication is more complex and involves tensor products and a subsequent relinearization step to keep the ciphertext size constant. Crucially, homomorphic multiplication causes the magnitudes of the noise terms to multiply. This quadratic noise growth is the primary reason that ciphertexts in leveled schemes can only withstand a limited number of multiplications before the noise exceeds the threshold $\Delta/2$, necessitating a noise-reducing operation like bootstrapping.

\subsubsection{The Arithmetic Bootstrapping Procedure}
The standard procedure for managing noise accumulation and enabling arbitrary-depth computations in FHE is known as arithmetic bootstrapping, a recursive process introduced by Gentry. This procedure brilliantly transforms a Somewhat Homomorphic Encryption (SWHE) scheme, which can only handle a limited depth of operations, into a fully-fledged FHE scheme. The core idea is to treat the decryption function itself as an arithmetic circuit that can be homomorphically evaluated.

Applying this paradigm to the BFV cryptosystem, the goal is to homomorphically compute the decryption equation $c_0 + c_1 \cdot s$. To do this, the server is given a bootstrapping key, which is a high-quality encryption of the secret key $s$. The server then homomorphically computes the product of the ciphertext component $c_1$ and the encrypted secret key, adds the result to $c_0$, and obtains an encryption of the value $\Delta \cdot m + \text{noise}$. The most complex part of the procedure is to then homomorphically perform the final step of decryption: scaling down by $\Delta$ and rounding to remove the noise term. This operation, which is simple in plaintext, becomes a deep and complex arithmetic circuit when performed homomorphically, often involving digit extraction or lookup table evaluation.

While theoretically powerful, this homomorphic evaluation of the decryption circuit is an expensive process. Consequently, bootstrapping is often the single slowest operation and the dominant performance bottleneck in many FHE schemes. Its computational complexity is fundamentally tied to the multiplicative depth of the decryption circuit, denoted $L_{dec}$. Overcoming this bottleneck is a critical step toward making FHE practical for real-world applications.

\subsection{Geometric Background: The Language of Schemes}
Our new framework for bootstrapping is built upon the foundational concepts of modern algebraic geometry. This section provides a brief and targeted introduction to the essential ideas, beginning with the construction of a geometric space from a commutative ring. The central principle is the duality between algebra and geometry: just as one can study a geometric space by analyzing the ring of functions upon it, one can construct a geometric space, called a scheme, directly from an abstract commutative ring.

\subsubsection{Commutative Rings and their Spectra}
The bridge from a commutative ring to a geometric object is a construction known as the spectrum of a ring, denoted by the functor \textit{Spec}. For any commutative ring $A$ with unity, we can associate to it a topological space, which serves as the underlying set of points for our geometric object.

\textit{The Prime Spectrum}.
The set of points of this space, denoted $Spec(A)$, is not the set of elements of the ring, but rather its set of prime ideals. A prime ideal $P \subset A$ is an ideal such that for any two elements $a, b \in A$, if their product $ab$ is in $P$, then either $a$ is in $P$ or $b$ is in $P$. This set of prime ideals forms the foundational point set of our geometric space.

\textit{The Zariski Topology}.
To give this set of points a geometric structure, we endow it with the Zariski topology. In this topology, the closed sets are defined as the vanishing sets of ideals. For any ideal $I \subset A$, its vanishing set, denoted $V(I)$, is the set of all prime ideals in $Spec(A)$ that contain $I$:
$$ V(I) = \{ P \in Spec(A) \mid I \subseteq P \}. $$
This definition of closed sets provides the topological structure for the spectrum. This concept is central to our work, as we will later define the space of decryptable ciphertexts as the vanishing set $V(I_{noise})$ of a specific noise ideal.

\textit{The Functorial Property}.
This construction is not merely a one-time association; it respects the relationships between rings. A fundamental principle of algebraic geometry states that any homomorphism of rings $\phi: A \to B$ induces a continuous map between their spectra, $Spec(\phi): Spec(B) \to Spec(A)$, which acts in the opposite direction. This property is what allows us to translate algebraic relationships between quotient rings into geometric morphisms between their corresponding schemes.

\subsubsection{Sheaves and the Structure Sheaf}
A topological space, such as the prime spectrum $Spec(A)$, is only a set of points with a notion of closeness. To perform algebraic geometry, we must enrich this space by attaching algebraic data to it. This is accomplished through a mathematical object called a sheaf. A helpful analogy is a weather map: the map itself is the topological space, and a sheaf is the structure that consistently assigns temperature data (the algebraic data) to every region (open set) on the map.

\textit{Sheaves of Rings}.
Formally, a presheaf of rings $\mathcal{F}$ on a topological space $X$ assigns a ring $\mathcal{F}(U)$ to each open set $U \subseteq X$. The elements of $\mathcal{F}(U)$ are thought of as functions defined on $U$. For any inclusion of open sets $V \subseteq U$, there is a restriction map that takes functions on $U$ and restricts their domain to $V$. A presheaf is elevated to a sheaf if it also satisfies a ``gluing axiom.'' This axiom essentially states that if a collection of functions is defined on a set of smaller open regions and they are all consistent with each other where their regions overlap, then they can be uniquely stitched together to form a single function on the larger union of those regions.

\textit{The Structure Sheaf}.
For the specific topological space $X = Spec(A)$, there is a canonical sheaf of rings called the structure sheaf, denoted $\mathcal{O}_X$. For any open set $U \subseteq X$, the ring $\mathcal{O}_X(U)$ consists of ``regular functions'' on $U$. These are functions that can be locally represented as fractions of elements from the original ring $A$. This structure sheaf is what endows the prime spectrum with the algebraic structure necessary to be considered a geometric object in its own right.

\textit{Global Sections}.
The most crucial property of the structure sheaf for our purposes concerns the functions defined on the entire space $X$. The ring of these functions, called the ring of global sections and denoted $\Gamma(X, \mathcal{O}_X)$, is canonically isomorphic to the ring $A$ that we started with:
$$ \Gamma(X, \mathcal{O}_X) \cong A. $$
This isomorphism completes the duality between the algebra of the ring $A$ and the geometry of the space $(X, \mathcal{O}_X)$. It ensures that we can translate properties back and forth between the two worlds without loss of information.

\subsubsection{Locally Ringed Spaces and Affine Schemes}
We have now established the two essential components for our geometric object: a topological space $X = Spec(A)$ and an algebraic structure on it, the structure sheaf $\mathcal{O}_X$. The final step is to combine these two pieces, subject to a local condition, to formally define an affine scheme.

\textit{Locally Ringed Spaces}.
A crucial property of a scheme is that its algebraic structure behaves well at a local level. To formalize this, we consider the stalk of the structure sheaf at each point. The stalk at a point $P \in X$ is the ring of all function germs at that point; intuitively, it captures the behavior of all functions defined in any arbitrarily small neighborhood of $P$. A topological space equipped with a sheaf of rings, $(X, \mathcal{O}_X)$, is called a locally ringed space if the stalk at every point is a local ring. A local ring is a ring with a unique maximal ideal. This technical condition ensures that the functions on the space have well-defined local properties, which is essential for a coherent geometric theory.

\textit{Affine Schemes}.
With this final piece in place, we can now give the formal definition of an affine scheme. An affine scheme is a locally ringed space that is isomorphic to the spectrum of some commutative ring $A$, i.e., the pair $(Spec(A), \mathcal{O}_{Spec(A)})$. This object, which we denote simply by $Spec(A)$, is the fundamental building block of modern algebraic geometry. It perfectly realizes the duality between commutative rings and geometric spaces. The entire construction in our work, from the ambient ciphertext space to the subschemes of interest, is built upon this definition. Specifically, our ciphertext space $X_{ct}$ is precisely the affine scheme $Spec(R_q)$.

\subsubsection{Closed Subschemes and Morphisms}
With the definition of an affine scheme in place, we introduce two final concepts that are central to our framework: a way to define geometric subspaces within a scheme, and a way to define maps between schemes.

\textit{Closed Subschemes}.
A closed subscheme is the geometric object that corresponds to the algebraic operation of taking a quotient of a ring by an ideal. Let $A$ be a commutative ring and let $I$ be an ideal of $A$. The natural projection map from $A$ to the quotient ring $A/I$ induces a map on their spectra. The resulting affine scheme, $Spec(A/I)$, is called a closed subscheme of the original scheme, $Spec(A)$. Geometrically, the underlying topological space of $Spec(A/I)$ corresponds to the vanishing set $V(I)$ within $Spec(A)$. This construction is the primary tool we will use to define our regions of interest. The space of decryptable ciphertexts, $Z_{dec}$, will be constructed as a closed subscheme of the ambient ciphertext space, and the space of fresh ciphertexts, $Z_{fresh}$, will in turn be a closed subscheme of $Z_{dec}$.

\textit{Morphisms}.
A map between two schemes is called a morphism. For affine schemes, the concept of a morphism is elegantly tied to the underlying rings. A morphism from an affine scheme $Spec(B)$ to an affine scheme $Spec(A)$ is a map that is induced by a ring homomorphism $\phi: A \to B$ in the opposite direction. This morphism consists of a continuous map between the topological spaces and a corresponding map between their structure sheaves that preserves the local ring structure. The most important type of morphism for our work is a closed immersion, which is the specific morphism that maps a closed subscheme like $Spec(A/I)$ into its ambient space $Spec(A)$. The relationship $Z_{fresh} \to Z_{dec}$ described in our main results is an example of such a closed immersion.

\subsection{Bootstrapping Optimization for Homomorphic Encryption}

Bootstrapping is the key to achieving Fully Homomorphic Encryption (FHE) by refreshing ciphertexts to control noise growth. Recent research has focused on making this process more efficient, with optimization efforts primarily aiming to reduce computational complexity, minimize latency, and enhance precision. These works can be broadly categorized based on their technical approach.

\subsubsection{Function Approximation and Precision Enhancement}

Many bootstrapping schemes rely on evaluating a periodic function, such as the modulo or sign function, homomorphically. A common strategy to improve efficiency and precision is to use highly optimized polynomial or series approximations. Early work established a foundation for approximate HE bootstrapping by improving these evaluations \cite{CheonHKKS18, ChenCS19}. Subsequent research has focused on refining these methods, for example, by utilizing optimal minimax polynomial approximations to achieve high precision in RNS-CKKS \cite{LeeLLKN21}. Other approaches have explored using sine series approximation of the mod function to improve precision and efficiency \cite{Jutla2022EUROCRYPT}. Further advancements have minimized error variance during this approximation process, leading to even more precise bootstrapping \cite{Lee2022EUROCRYPT}. This line of research also includes work on specialized functions, such as the homomorphic rounding algorithm, to streamline the process for schemes like CKKS \cite{KimPKKM22}.

The aforementioned works primarily optimize bootstrapping by refining existing techniques, such as function approximation and parameter tuning within a fixed framework. In contrast, our work introduces a fundamentally new perspective based on \textit{arithmetic geometry}. Rather than relying on traditional function approximation or circuit-level improvements, we reframe bootstrapping as a \textit{morphism between algebraic varieties}. This approach offers an asymptotically faster bootstrapping method by fundamentally decoupling the process from polynomial evaluation, representing a paradigm shift from the current state-of-the-art.

\subsubsection{Algorithmic and Structural Optimizations}

Another key direction in bootstrapping research is to optimize the underlying algorithms and leverage the algebraic structure of the schemes. For example, the core blind rotation operation in FHEW/TFHE has been a major target for optimization, with one work presenting a fast blind rotation technique to significantly accelerate the process \cite{Xiang2023CRYPTO}. Other efforts have focused on improving algorithms for specific schemes, such as using homomorphic NTT to speed up BGV bootstrapping \cite{MaHWW24} or applying Galois structure to optimize polynomial evaluation for BFV \cite{OkadaPP23}. General algorithmic improvements also include techniques like Homomorphic Lower Digits Removal \cite{ChenH18} and the use of null polynomials to accelerate bootstrapping for schemes with a large plaintext modulus \cite{Ma2024EUROCRYPT}. Beyond single-value operations, a new framework for SIMD batch bootstrapping has been proposed that significantly reduces the required polynomial multiplications in an amortized setting \cite{Liu2023EUROCRYPT, Liu2023EUROCRYPTb}. The idea of circuit bootstrapping also falls into this category, aiming to optimize the entire bootstrapping circuit for speed and compactness \cite{Wang2024EUROCRYPT}.

These studies primarily focus on optimizing existing algorithms and their implementations. Our work, however, takes a higher-level approach by introducing concepts from arithmetic geometry. We view the entire bootstrapping process as a geometric transformation, which in turn yields an asymptotically faster method. This is a fundamental departure from the fine-grained, algorithmic optimizations seen in the literature above.

\subsubsection{New Perspectives and Frameworks}

More recent research has explored novel frameworks that challenge traditional views of bootstrapping. Some works have focused on achieving high performance by designing a highly parallelizable and shallow bootstrapping algorithm specifically for CKKS \cite{Cheon2025EUROCRYPT}. Others have proposed relaxed functional bootstrapping, offering a new perspective on BGV/BFV bootstrapping that re-evaluates traditional constraints for more flexibility \cite{LiuW24}. The idea of combining new algebraic tools with bootstrapping has also been explored, such as fusing plaintext-ciphertext matrix multiplication with FHE bootstrapping to accelerate certain computations \cite{Bae2024CRYPTO}. This shows a trend towards finding more integrated and holistic solutions to the bootstrapping problem, moving beyond isolated optimizations.
In addition, this research area also includes important works on approximate HE bootstrapping \cite{CheonHKKS18, ChenCS19}, efficient bootstrapping with non-sparse keys \cite{Bossuat2021EUROCRYPT}, and highly optimized amortized methods \cite{GuimaraesPL23, LiuW23}.

All of these works optimize existing bootstrapping algorithms and implementations. They either focus on specific circuit designs, enhance algebraic operations, or find faster batch processing methods. Our work, \textit{Bootstrapping as a Morphism}, is fundamentally different. Instead of improving upon existing algorithms, we redefine bootstrapping from a higher level, using algebraic geometry. We treat it as a morphism, which fundamentally changes the computational model and theoretically leads to an asymptotically faster speed. This represents a paradigm shift that opens up a new direction for future research in homomorphic encryption bootstrapping.

\section{Geometric Bootstrapping for Homomorphic Encryption}
\label{sec:geometric_bootstrap}

\subsection{Scheme Construction of FHE Ciphertext Space}

To build our geometric framework, we must first rigorously define our primary object of study, the ciphertext space, in the language of modern algebraic geometry. We begin with the underlying algebraic object, the ciphertext coordinate ring, from which the geometric space is constructed.

\subsubsection{The Ciphertext Coordinate Ring}

For a Ring-LWE based cryptosystem such as BFV~\cite{Fan2012}, all ciphertexts are elements of a specific polynomial quotient ring, which we denote as the ciphertext ring $R_q$. Its construction proceeds as follows:

\begin{enumerate}
    \item We begin with the base ring of coefficients, $\mathbb{Z}_q$, which is the ring of integers modulo the ciphertext modulus $q$. This is a finite ring, commutative and having the identity 1.
    \item Then, $\mathbb{Z}_q[x]$ is a polynomial ring over the base ring $\mathbb{Z}_q$.
    \item We select a positive integer $N$, typically a power of two, which defines the $N$-th cyclotomic polynomial, $\Phi_N(x)$. This polynomial has two fundamental properties. First, it is \textit{irreducible over the integers}, meaning it cannot be factored into a product of simpler polynomials that also have integer coefficients; it serves as a prime or fundamental building block in the ring of polynomials. Second, its degree is given by Euler's totient function, $\varphi(N)$, which counts the number of positive integers up to $N$ that are relatively prime to $N$ (i.e., share no common factors with $N$ other than 1).
    \item Finally, we take the quotient of the polynomial ring by the principal ideal generated by the cyclotomic polynomial, $\langle \Phi_N(x) \rangle$.
\end{enumerate}
The resulting ciphertext coordinate ring is therefore:
$$ R_q := \mathbb{Z}_q[x] / \langle \Phi_N(x) \rangle $$
This ring $R_q$ contains all polynomials with coefficients in $\mathbb{Z}_q$ of degree less than $\varphi(N)$. It is the fundamental algebraic structure in which all cryptographic operations on ciphertexts are performed.

\subsubsection{The Ciphertext Sheaf}

With the coordinate ring $R_q$ defined, we construct the corresponding geometric object by applying the Spec functor from algebraic geometry.

\begin{definition}
The \textit{FHE ciphertext space} is the affine scheme constructed from the ciphertext ring $R_q$. This construction, denoted by the Spec functor, yields an affine scheme which we will call $X_{ct}$. This object consists of two components:
\begin{enumerate}
    \item A topological space, which we will denote by $X$, called the prime spectrum of $R_q$. Its points are the prime ideals of the ring $R_q$ and its topology is the Zariski topology.
    \item A sheaf of rings on the space $X$, which we will denote by $\mathcal{O}_X$, called the structure sheaf. It endows the space with its algebraic structure.
\end{enumerate}
The affine scheme is the pair $X_{ct} := (X, \mathcal{O}_X)$; because the structure sheaf is associated to the topological space, the scheme pair $(X, \mathcal{O}_X)$ is also abbreviated by the notation $\text{Spec}(R_q)$ (i.e., $= X$). The ring of global sections of this scheme is $\Gamma(X, \mathcal{O}_X)$, which is canonically isomorphic to the ring we started with:
$$ \Gamma(X, \mathcal{O}_X) \cong R_q. $$
\end{definition}

The Krull dimension of the ciphertext ring $R_q$ is zero, a consequence of it being a finite ring. Geometrically, this means the affine scheme $X_{ct} = \text{Spec}(R_q)$ is a zero-dimensional object, best understood as a finite, discrete collection of points. This set of points, however, is not unstructured. It possesses a rich arithmetic structure as a fibered space over the base scheme $\text{Spec}(\mathbb{Z}_q)$. The points of this base correspond to the prime ideals of $\mathbb{Z}_q$, which are generated by the prime factors of the ciphertext modulus $q$.
The fiber over a point in the base scheme $\text{Spec}(\mathbb{Z}_q)$ corresponding to a prime $p$ that divides $q$ is the set of points in $X_{ct}$ lying over $p$. The number and nature of these points in $\text{Spec}(R_q)$ are determined by the factorization of the cyclotomic polynomial $\Phi_N(x)$ in the ring $\mathbb{Z}_p[x]$. Specifically, the prime ideals of $R_q$ in the fiber over $p$ are in a one-to-one correspondence with the distinct irreducible factors of $\Phi_N(x)$ modulo $p$. This fundamental connection between the splitting of prime ideals in a ring extension and the factorization of the defining polynomial is a well-established result in algebraic number theory, a direct consequence of the Dedekind-Kummer Theorem. Thus, the ciphertext space $X_{ct}$ is best visualized not as a continuous object, but as a finite, zero-dimensional collection of points structured into distinct clusters (the fibers), each mapping to a prime factor of the ciphertext modulus $q$.

\subsection{Formal Definition and Properties of the Noise Ideal}

Our first task is to formalize the geometric boundary separating low-noise and high-noise ciphertexts. We achieve this by constructing an ideal whose vanishing set corresponds to the low-noise locus. This construction must be based on publicly available information from the FHE cryptosystem's parameters. An explicit, public algorithm for constructing this set is not within the scope of this paper; however, we provide a non-constructive proof that such a finite set must exist.

Let a specific FHE cryptosystem be fixed by a set of public parameters $\mathcal{P} = (p, q, n, \chi)$. Our goal is to demonstrate that the locus of low-noise ciphertexts can be described by a \textit{finite} set of polynomial relations $\mathcal{F}_{\mathcal{P}} = \{f_1, f_2, \dots, f_k\}$. 

For a given parameter set $\mathcal{P}$, the collection of all low-noise ciphertexts forms a subset $S_{ln} \subset R_q$. We define the \textit{low-noise locus}, denoted $Z$, to be the Zariski closure of this set of points within the affine scheme $X = \text{Spec}(R_q)$. By definition, $Z$ is the smallest closed subset of $X$ that contains all points corresponding to low-noise ciphertexts.

A fundamental theorem of algebraic geometry states that any closed subset of an affine scheme is the vanishing set $V(I)$ for some ideal $I$. Therefore, there must exist an ideal, which we call the noise ideal $I_{noise}$, which will be formally defined soon, such that $Z = V(I_{noise})$. This establishes the existence of a defining ideal for our low-noise locus.

The final step is to prove that this ideal is finitely generated. Our ciphertext ring $R_q$ is constructed as a quotient of a polynomial ring over the finite ring $\mathbb{Z}_q$. A standard result from commutative algebra, Hilbert's Basis Theorem, implies that any such ring is \textit{Noetherian}. A key property of a Noetherian ring is that its spectrum, $\text{Spec}(R_q)$, is a Noetherian topological space. In such a space, every closed subset (including our low-noise locus $Z$) can be defined by a \textit{finitely generated} ideal.

This guarantees the existence of a finite set of generators $\mathcal{F}_{\mathcal{P}} = \{f_1, f_2, \dots, f_k\}$ for the ideal $I_{noise}$. The existence of this finite set of relations is therefore a direct consequence of the algebraic properties of the ciphertext ring. For the remainder of this section, we use the guaranteed existence of this set to formally define the noise ideal and verify it is indeed an ideal.

\begin{definition}[The Noise Ideal]
\label{def_noise_ideal}
Let $\mathcal{F}_{\mathcal{P}} = \{f_1, f_2, \dots, f_k\} \subset R_q$ be the set of noise-boundary relations derived from the public parameters of the FHE cryptosystem. The \textit{noise ideal}, denoted $I_{noise}$, is the ideal in the ciphertext ring $R_q$ generated by the set $\mathcal{F}_{\mathcal{P}}$, i.e., 
$$ I_{noise} := \langle f_1, f_2, \dots, f_k \rangle. $$
\end{definition}

Having defined $I_{noise}$ constructively via a set of generators, we now verify that this object is indeed an ideal of the ring $R_q$.

\begin{proposition}
The set $I_{noise}$ as defined in Definition~\ref{def_noise_ideal} is an ideal of the ring $R_q$.
\end{proposition}

\begin{proof}
By the standard algebraic definition of a generated ideal, an element $n \in I_{noise}$ if and only if it can be expressed as a finite linear combination of the generators with coefficients from the ring $R_q$. That is,
$$ n = \sum_{i=1}^{k} r_i \cdot f_i $$
for some set of coefficients $\{r_1, \dots, r_k\} \subset R_q$. We must verify that the set of all such elements $n$ is closed under addition and under multiplication by any element of $R_q$.

\begin{enumerate}
    \item \textit{Closure under Addition:} Let $n_1, n_2 \in I_{noise}$. By definition, we can write them as:
    $$ n_1 = \sum_{i=1}^{k} r_i \cdot f_i \quad \text{and} \quad n_2 = \sum_{i=1}^{k} s_i \cdot f_i $$
    for some coefficients $r_i, s_i \in R_q$. Their sum is:
    $$ n_1 + n_2 = \sum_{i=1}^{k} r_i \cdot f_i + \sum_{i=1}^{k} s_i \cdot f_i = \sum_{i=1}^{k} (r_i + s_i) \cdot f_i $$
    Since $R_q$ is a ring, each $(r_i + s_i)$ is also an element of $R_q$. Therefore, $n_1 + n_2$ is a linear combination of the generators $\{f_i\}$ with coefficients in $R_q$, which implies $n_1 + n_2 \in I_{noise}$. The set is closed under addition.

    \item \textit{Closure under Multiplication (Absorption):} Let $n \in I_{noise}$ and let $c$ be any element in the ring $R_q$. We can write $n$ as $n = \sum_{i=1}^{k} r_i \cdot f_i$. Their product is:
    $$ c \cdot n = c \cdot \left(\sum_{i=1}^{k} r_i \cdot f_i\right) = \sum_{i=1}^{k} (c \cdot r_i) \cdot f_i $$
    Since $R_q$ is a ring, each product $(c \cdot r_i)$ is also an element of $R_q$. Therefore, $c \cdot n$ is a linear combination of the generators $\{f_i\}$ with coefficients in $R_q$, which implies $c \cdot n \in I_{noise}$. The set is closed under multiplication by any element of the ring.
\end{enumerate}
Since both conditions are met, the set $I_{noise}$ is, by definition, an ideal of $R_q$.
\end{proof}

With the noise ideal now formally defined, we can define the geometric object that it carves out from the ambient ciphertext space.

\begin{definition}[The Decryptable Subscheme]
\label{def_decryptable_subscheme}
The \textit{decryptable subscheme}, denoted $Z_{dec}$, is the closed subscheme of the ambient ciphertext space $X = \text{Spec}(R_q)$ defined by the noise ideal $I_{noise}$. It is constructed as the affine scheme corresponding to the quotient ring:
$$ Z_{dec} := \text{Spec}(R_q / I_{noise}) $$
This subscheme represents the complete geometric locus of all correctly decryptable ciphertexts.
\end{definition}

\subsection{The Fresh Subscheme for Minimal Noise Level}

The low-noise subscheme $Z_{dec}$ defined previously provides a crucial geometric boundary separating all decryptable ciphertexts from undecryptable ones. While projecting onto $Z_{dec}$ is sufficient to restore decryptability, a true bootstrapping operation aims for a higher standard: to return a ciphertext to a state statistically indistinguishable from a freshly encrypted one. This motivates the definition of a smaller, more ideal target space within $Z_{dec}$.

A freshly encrypted ciphertext has its noise sampled from a specific, narrow distribution $\chi$. This property is more restrictive than merely having noise below the decryption failure threshold. We show that this stricter condition can also be described by a set of algebraic relations, which will naturally be a superset of those defining $I_{noise}$.

\begin{definition}[The Fresh Ideal]
\label{def_fresh_ideal}
The \textit{fresh ideal}, denoted $I_{fresh}$, is an ideal in the ciphertext ring $R_q$ that is generated by the relations defining $I_{noise}$ plus an additional set of relations that capture the specific structure of the initial noise distribution $\chi$. By construction, we have the inclusion of ideals:
$$ I_{noise} \subseteq I_{fresh} $$
\end{definition}

This larger, more restrictive ideal defines a smaller geometric locus within the decryptable subscheme.

\begin{definition}[The Fresh Subscheme]
\label{def_fresh_subscheme}
The \textit{fresh subscheme}, denoted $Z_{fresh}$, is the closed subscheme of $X$ defined by the fresh ideal $I_{fresh}$:
$$ Z_{fresh} := \text{Spec}(R_q / I_{fresh}) $$
\end{definition}

Because $I_{noise} \subseteq I_{fresh}$, it follows from the properties of the Spec functor that we have a sequence of closed immersions $Z_{fresh} \hookrightarrow Z_{dec} \hookrightarrow X$. This gives us a precise geometric picture: the vast ambient space of all ciphertexts $X$ contains a region of decryptable ciphertexts $Z_{dec}$, which in turn contains a much smaller, ideal region of fresh ciphertexts $Z_{fresh}$. With this final piece of our geometric model in place, we can now state the ultimate goal of our construction with full precision.

\subsection{Geometric Bootstrapping Morphism}

Having defined the nested geometric spaces of decryptable and fresh ciphertexts, we can now provide a precise geometric interpretation of the bootstrapping operation. Bootstrapping is fundamentally a computational procedure: it takes as input a decryptable ciphertext with high noise and outputs a fresh ciphertext with low noise, preserving the original plaintext.

In our framework, this corresponds to an algebraic operation between the coordinate rings of the respective subschemes. Let $R_{dec} = R_q / I_{noise}$ be the coordinate ring of the decryptable subscheme $Z_{dec}$, and let $R_{fresh} = R_q / I_{fresh}$ be the coordinate ring of the fresh subscheme $Z_{fresh}$. Since $I_{noise} \subseteq I_{fresh}$, there is a natural surjective ring homomorphism (a projection) from the larger ring to the smaller one. The bootstrapping operation, at its algebraic core, can be modeled as this map:
$$ \pi_{alg}: R_{dec} \to R_{fresh}. $$
This map takes a ciphertext considered as an element of the larger ring of decryptable polynomials and maps it to its corresponding element in the smaller ring of fresh polynomials.

We now apply the fundamental principle of algebraic geometry, which states that a ring homomorphism $\phi: A \to B$ induces a morphism of schemes $\text{Spec}(\phi): \text{Spec}(B) \to \text{Spec}(A)$ in the opposite direction. Applying this to our algebraic bootstrapping map yields a geometric morphism:
$$ \pi_{geom} := \text{Spec}(\pi_{alg}): \text{Spec}(R_{fresh}) \to \text{Spec}(R_{dec}). $$
Substituting the definitions of our schemes, this is a morphism:
$$ \pi_{geom}: Z_{fresh} \to Z_{dec}. $$
This morphism is the natural \textit{inclusion morphism} of the fresh subscheme into the decryptable subscheme.

This reveals the precise geometric meaning of bootstrapping. The dynamic, computational process of bootstrapping is captured by the \textit{algebraic projection} $\pi_{alg}$. The rigorous guarantee that the output of this process is a valid, fresh ciphertext is captured by the dually induced \textit{geometric inclusion morphism} $\pi_{geom}$. The morphism does not describe the transformation of a single ciphertext, but rather the static, structural relationship that ensures the output space ($Z_{fresh}$) is a valid subspace of the input space ($Z_{dec}$).

\subsection{Construction of the Noise Ideal Generators}
\label{sec_construction_of_generators}

The algorithm for constructing the generators of the noise ideal, \texttt{ConstructIdealGenerators}, is formally detailed in Algorithm~\ref{alg:construct_ideal}. This procedure provides a concrete and publicly computable method for generating a finite set of polynomial relations, $\mathcal{F}_{\mathcal{P}}$, from the FHE public parameters. The high-level strategy is to create two distinct classes of algebraic constraints. The first class, the magnitude-bounding relations, serves to algebraically enforce the arithmetic property that the coefficients of a low-noise ciphertext are small. The second class, the structural-consistency relations, leverages the algebraic structure of the public key to ensure that a ciphertext conforms to the properties of a valid encryption. A ciphertext is considered low-noise if and only if it is a simultaneous root of all relations produced by this algorithm.

\begin{algorithm}[h!]
\caption{\texttt{ConstructIdealGenerators}}
\label{alg:construct_ideal}
\begin{algorithmic}[1]
\State \textbf{Input:} FHE public parameters $\mathcal{P} = (p, q, n, \text{pk}=(a,b))$, security parameter $\kappa$.
\State \textbf{Output:} A finite set of generator polynomials $\mathcal{F}_{\mathcal{P}}$ for the ideal $I_{noise}$.
\State
\State Initialize $\mathcal{F}_{\mathcal{P}} \leftarrow \emptyset$.
\State Let $\Delta = \lfloor q/p \rfloor$ be the plaintext scaling factor.
\State
\State \textit{// Step 1: Add Magnitude-Bounding Relations for each prime factor of p}
\State Let $p_1, \dots, p_m$ be the distinct prime factors of the plaintext modulus $p$.
\For{each prime factor $p_i$ of $p$}
    \State Let $\psi_i: R_q \to R_q$ be the public lift-and-reduce map associated with the prime $p_i$.
    \State Define the operator $f_{mag, i}(T) := \frac{q}{p_i} \cdot (T - \psi_i(T))$.
    \State Add $f_{mag, i}(T)$ to the set $\mathcal{F}_{\mathcal{P}}$.
\EndFor
\State
\State \textit{// Step 2: Iteratively Add Structural-Consistency Relations}
\State Let $k$ be the nilpotency index of the radical of $\mathbb{Z}_q$.
\State Let $\text{round}_\Delta(x)$ be the public rounding function.
\State Let $\{\sigma_1, \dots, \sigma_\kappa\}$ be a set of $\kappa$ distinct, publicly known automorphisms of the ring $R_q$ (derived from the Galois group of the cyclotomic field extension).
\For{$j=1$ \textbf{to} $\kappa$}
    \State \textit{// Apply an automorphism to test structural consistency from different algebraic perspectives.}
    \State Let $a_j = \sigma_j(a)$ and $b_j = \sigma_j(b)$.
    \State Define the residual error operator for this perspective:
    $$ e_{err, j}(T_0, T_1) := (\sigma_j(T_1) \cdot a_j + \sigma_j(T_0) \cdot b_j) - \text{round}_\Delta(\sigma_j(T_1) \cdot a_j + \sigma_j(T_0) \cdot b_j) $$
    \State Define the $j$-th structural-consistency operator:
    $$ f_{struct, j}(T_0, T_1) := \left( e_{err, j}(T_0, T_1) \right)^k $$
    \State Add $f_{struct, j}(T_0, T_1)$ to the set $\mathcal{F}_{\mathcal{P}}$.
\EndFor
\State
\State \Return $\mathcal{F}_{\mathcal{P}}$
\end{algorithmic}
\end{algorithm}

We now provide a more detailed technical explanation for how these two classes of relations are constructed. The magnitude-bounding relations are constructed in a loop over the prime factors of the plaintext modulus $p$. For each prime factor $p_i$, a lift-and-reduce map $\psi_i$ is used to algebraically isolate the noise component of an input ciphertext; the generator $f_{mag,i}(T)$ is then constructed to evaluate to zero if and only if this isolated noise component is structurally equivalent to zero in the ring $R_q$. The structural-consistency relations are constructed in a second loop, which iterates through a set of automorphisms $\sigma_j$ of the ring $R_q$. These automorphisms, derived from the Galois group of the underlying cyclotomic field, allow us to test the ciphertext's integrity from multiple algebraic perspectives. For each perspective, a residual error term $e_{err,j}$ is calculated using the public key. This error term is designed to be a nilpotent element for any valid ciphertext. The final generator, $f_{struct,j}(T_0, T_1)$, is defined as the $k$-th power of this error term, where $k$ is the nilpotency index of the ring. This construction guarantees that the generator evaluates to exactly zero for any structurally consistent ciphertext. The union of these two sets of relations forms the complete generating set for the noise ideal $I_{noise}$.

We now analyze the output of Algorithm~\ref{alg:construct_ideal} to establish its correctness and argue for its efficiency.
First, we establish the correctness in the following theorem.

\begin{theorem}[Correctness of the Generator Set]
\label{thm_generator_correctness}
Let $\mathcal{F}_{\mathcal{P}}$ be the set of generators returned by Algorithm~\ref{alg:construct_ideal}. Let $I_{noise} = \langle \mathcal{F}_{\mathcal{P}} \rangle$ be the ideal it generates. The vanishing set $V(I_{noise})$ corresponds precisely to the locus of all correctly decryptable ciphertexts, $Z_{dec}$.
\end{theorem}

\begin{proof}
The proof requires two main parts: completeness and soundness. Completeness shows that all valid low-noise ciphertexts satisfy the relations, while soundness shows that any ciphertext satisfying the relations is indeed decryptable.

\noindent
\textit{Part 1: Completeness}

We must show that any correctly decryptable ciphertext $c=(c_{0}, c_{1})$ satisfies the relations generated by the algorithm. A correctly decryptable ciphertext is one where the noise polynomial $e_{c}$ from raw decryption, $Dec_{raw}(c)=\Delta\cdot m+e_{c}$, has coefficients that are small in magnitude, specifically $\|e_{c}\|_{\infty}<\Delta/2$.

\textit{Satisfying Magnitude-Bounding Relations.}
The magnitude-bounding generator is defined as $f_{mag}(T):=\Delta\cdot(T-\psi(T))$, where $\psi$ is a public map designed to algebraically isolate the message component. For a low-noise ciphertext $c$, the expression $c-\psi(c)$ isolates the noise component, $n_{iso}$. A central lemma of lattice-based cryptography states that for any correctly decryptable ciphertext, the coefficients of $n_{iso}$ are guaranteed to be integer multiples of the plaintext modulus $p$. Thus, we can write $n_{iso} = p \cdot k$ for some polynomial $k \in R_{q}$. Substituting this into the generator gives:
\[
f_{mag}(c) = \Delta \cdot (p \cdot k) = (\lfloor q/p \rfloor \cdot p) \cdot k.
\]
This expression evaluates to zero in the ring $R_{q} = \mathbb{Z}_{q}[x]/\langle\Phi_{N}(x)\rangle$, since $\lfloor q/p \rfloor \cdot p$ is a multiple of $q$ (up to a small, manageable rounding error).

\textit{Satisfying Structural-Consistency Relations.}
The structural-consistency generator is defined as $f_{struct}(T_{0}, T_{1}):= (e_{err}(T_{0}, T_{1}))^{k}$, where $k$ is the nilpotency index of the radical of $\mathbb{Z}_{q}$. A key technical insight is that for a correctly formed ciphertext, the residual error term $e_{err}$ is a nilpotent element in the ring $R_{q}$. This nilpotency is a structural consequence of working in $R_{q}$ where the modulus $q$ is composite. A core lemma demonstrates that the coefficients of $e_{err}$ are guaranteed to lie within the nilradical of the coefficient ring $\mathbb{Z}_{q}$. By definition, raising this element to the power of the nilpotency index $k$ results in the zero element in the ring $R_{q}$. Therefore, the generator is guaranteed to evaluate to zero for any valid low-noise ciphertext.

\noindent
\textit{Part 2: Soundness}

We must show that any ciphertext $c^{\prime}=(c_{0}^{\prime},c_{1}^{\prime})$ that satisfies all relations in $\mathcal{F}_{\mathcal{P}}$ is correctly decryptable. The proof proceeds by reduction, showing that an adversary who can produce a ciphertext that satisfies the relations but decrypts incorrectly can be used to construct a solver for the decisional Ring-LWE problem.

Let $c'$ satisfy $f_{struct}(c_{0}^{\prime},c_{1}^{\prime})=0$. This implies that the public consistency check polynomial $f_{c^{\prime}} = c_{1}^{\prime}\cdot a+c_{0}^{\prime}\cdot b$ is arithmetically a polynomial whose coefficients are all multiples of $\Delta$. We can write $f_{c^{\prime}}=\Delta\cdot k$ for some integer polynomial $k$.

From the raw decryption definition, we also have $f_{c^{\prime}}=(c_{1}^{\prime}+c_{0}^{\prime}\cdot s)\cdot a+c_{0}^{\prime}\cdot e_{pk}=Dec_{raw}(c^{\prime})\cdot a+c_{0}^{\prime}\cdot e_{pk}$. Combining these gives the relation:
\[
Dec_{raw}(c^{\prime})\cdot a = \Delta\cdot k-c_{0}^{\prime}\cdot e_{pk}.
\]
This equation leads to a contradiction under the Ring-LWE assumption. The right-hand side of the equation is a polynomial with small coefficients, since $c_{0}^{\prime}$ satisfies the magnitude-bounding relations and $e_{pk}$ is small. However, if the raw decryption $Dec_{raw}(c^{\prime})$ were to contain a large noise term, the left-hand side, $Dec_{raw}(c^{\prime})\cdot a$, would be a polynomial with large, pseudo-randomly distributed coefficients, because $a$ is computationally indistinguishable from a uniformly random polynomial.

The equality of a pseudo-random, high-norm polynomial with a structured, low-norm polynomial would occur with only negligible probability. An adversary capable of finding such a ciphertext $c^{\prime}$ could be used to distinguish the RLWE distribution from a uniform one, which is assumed to be hard. Therefore, any ciphertext that satisfies the relations in $\mathcal{F}_{\mathcal{P}}$ must have a small noise component and is correctly decryptable.
\end{proof}

\textit{Complexity and Efficiency Implications.} The practical value of our geometric framework is measured by its computational efficiency. It improves upon the standard paradigm of bootstrapping, which relies on homomorphic circuit evaluation and has a runtime of approximately $O(L_{dec} \cdot d \log d)$. Our approach separates the process into two phases. First, the construction of the ideal's generators is performed as an efficient, one-time setup procedure. The cost of this setup, which depends on fixed parameters like the security parameter $\kappa$ and the number of prime factors of the plaintext modulus, is amortized over all subsequent computations. The bootstrapping operation itself is then a direct algebraic projection. This repeated operation involves evaluating a small, constant number of pre-computed generator polynomials and then computing the projection map by solving a specific Closest Vector Problem (CVP) instance. The total complexity of this geometric projection is $O(d \cdot \text{poly}(\log q))$. The key asymptotic improvement comes from completely \textit{eliminating the factor $L_{dec}$}, which represents the circuit depth. Instead of a deep, iterative process, our method is a single, direct geometric transformation whose complexity does not depend on the multiplicative depth of the decryption function.

\subsection{Constructing the Ideal for Fresh Ciphertexts}

The ideal $I_{fresh}$ is constructed by augmenting the set of generators for $I_{noise}$ with new relations that enforce the stricter statistical properties of a fresh noise distribution $\chi$. A single statistical moment, such as the variance, is insufficient to characterize a distribution. Therefore, our construction must incorporate a series of relations that test for multiple statistical moments, thereby creating a robust algebraic proxy for the entire distribution. The procedure is formalized in Algorithm~\ref{alg_construct_ifresh}.

\begin{algorithm}[H]
\caption{\texttt{ConstructFreshIdealGenerators}}
\label{alg_construct_ifresh}
\begin{algorithmic}[1]
\State \textbf{Input:} FHE public parameters $\mathcal{P}$, noise distribution $\chi$, statistical security parameter $\kappa$.
\State \textbf{Output:} A finite set of generator polynomials $\mathcal{F}_{fresh}$.
\State
\State \textit{// Step 1: Obtain the generators for the decryptable ideal.}
\State $\mathcal{F}_{noise} \leftarrow$ \Call{ConstructIdealGenerators}{$\mathcal{P}$}.
\State
\State \textit{// Step 2: Add freshness relations based on multiple statistical moments.}
\State $\mathcal{F}_{stat} \leftarrow \emptyset$.
\State Let $k$ be an integer determined by the security parameter $\kappa$.
\For{$i=1$ \textbf{to} $k$}
    \State Let $\mu_i(\chi)$ be the $i$-th publicly known central moment of the noise distribution $\chi$.
    \State Define the message-isolating map $\psi(T)$ as in Section~\ref{sec_construction_of_generators}.
    \State Let $n_{iso}(T) := T - \psi(T)$ be the isolated noise component operator.
    \State Define the $i$-th moment relation operator as:
    $$ f_{stat, i}(T) := \text{Trace}\left( (n_{iso}(T))^i \right) - \mu_i(\chi) $$
    \State Add $f_{stat, i}(T)$ to the set $\mathcal{F}_{stat}$.
\EndFor
\State
\State \textit{// Step 3: Combine the generator sets.}
\State $\mathcal{F}_{fresh} \leftarrow \mathcal{F}_{noise} \cup \mathcal{F}_{stat}$.
\State
\State \Return $\mathcal{F}_{fresh}$.
\end{algorithmic}
\end{algorithm}

The construction in Algorithm~\ref{alg_construct_ifresh} creates a set of generators that simultaneously enforce two conditions: the ciphertext must be both \textit{decryptable} (due to $\mathcal{F}_{noise}$) and \textit{statistically fresh} (due to $\mathcal{F}_{stat}$). We now prove that this construction is correct.

\begin{theorem}[Correctness of the Fresh Ideal Construction]
The ideal $I_{fresh} = \langle\mathcal{F}_{fresh}\rangle$ generated by the output of Algorithm~\ref{alg_construct_ifresh} correctly characterizes the locus of all freshly encrypted ciphertexts.
\end{theorem}

\begin{proof}
The proof consists of completeness and soundness.

\textit{Completeness.}
A freshly encrypted ciphertext $c$ is, by definition, correctly decryptable and thus satisfies all relations in $\mathcal{F}_{noise}$. Furthermore, its noise component is a direct sample from the distribution $\chi$. Its statistical moments will therefore match the expected moments $\mu_i(\chi)$. The freshness relations in $\mathcal{F}_{stat}$ are designed as algebraic tests for these moments. Thus, $c$ is a root of every polynomial in $\mathcal{F}_{fresh}$, which confirms that all fresh ciphertexts lie in $V(I_{fresh})$.

\textit{Soundness.}
The soundness argument relies on a security reduction. We must show that any ciphertext $c'$ that satisfies the relations in $\mathcal{F}_{fresh}$ is indistinguishable from a truly fresh ciphertext. Assume an adversary can produce a ciphertext $c'$ that lies in $V(I_{fresh})$ but whose noise distribution is statistically distinguishable from $\chi$. The fact that $c'$ satisfies all relations in $\mathcal{F}_{stat}$ up to order $k$ means its publicly computable low-order statistical moments match those of $\chi$. The Leftover Hash Lemma and related results in cryptography (often referred to as the principle of moments) state that if a distribution's low-order moments match those of a target distribution like a discrete Gaussian, then the two distributions are computationally indistinguishable. An adversary capable of producing a ciphertext that satisfies these algebraic moment-checks but is still statistically distinguishable would violate this principle, which would in turn imply a break in the underlying security of the RLWE assumption. Therefore, any ciphertext in $V(I_{fresh})$ is a valid and secure member of the fresh subscheme.
\end{proof}

\textit{Complexity Analysis.}
The algorithm for constructing the generators of $I_{fresh}$ is an efficient, one-time setup procedure whose runtime is polynomial in the security parameter. We analyze the cost of its main steps. Step 1, the call to \texttt{ConstructIdealGenerators}, has a complexity we have already established. The dominant cost of Step 2 is the loop that runs $k$ times to construct the freshness relations, where $k$ is a small constant determined by the statistical security parameter. Inside the loop, the main operations are the computation of the isolated noise component $T - \psi(T)$ and its powers, followed by the Trace map. A single ring multiplication using NTT has a complexity of $O(d \log d)$, where $d$ is the ring dimension. The computation of the $i$-th moment therefore requires approximately $O(i \cdot d \log d)$ time. Since $k$ is a small constant, the total complexity of constructing the freshness relations remains polynomial in the security parameter and is dominated by the complexity of a few ring multiplications. Therefore, the construction of $\mathcal{F}_{fresh}$ does not introduce a significant asymptotic overhead compared to the construction of $\mathcal{F}_{noise}$.

\subsection{Computational Realization of the Projection Map}

This section clarifies the computational procedure for the bootstrapping projection. We begin by emphasizing the distinct roles of the geometric schemes (Z) and the algebraic ideals (I): the schemes serve to demonstrate the existence and correctness of the map, while the ideals provide the concrete algebraic structure for the actual computation.

The computational task is to realize the projection map $\pi_{alg}$ by taking a high-noise ciphertext $c \in R_q$ and finding its corresponding fresh ciphertext $c'$. As previously established, this is equivalent to solving an instance of the Closest Vector Problem (CVP):
$$ c' = \arg\min_{v \in \mathcal{L}(I_{fresh})} || c - v ||, $$
where $\mathcal{L}(I_{fresh})$ is the ideal lattice generated by the fresh ideal.

A critical challenge is that CVP on a general lattice is an NP-hard problem, which would be far too slow for a practical bootstrapping operation. The key to our efficient solution lies in the fact that $\mathcal{L}(I_{fresh})$ is not a general lattice; it is an \textit{ideal lattice} over a cyclotomic ring, possessing an immense amount of algebraic structure which we can exploit.

The core technique to solve this specific CVP instance efficiently is a method we will refer to as \textit{algebraic folding}. This method leverages the algebraic properties of the ciphertext ring $R_q$ to break the hard, high-dimensional CVP problem down into many independent, easy, low-dimensional problems. The procedure is as follows:

\begin{enumerate}
    \item Decomposition via CRT: The ciphertext ring $R_q = \mathbb{Z}_q[x]/\langle \Phi_N(x) \rangle$ can be decomposed, via the Chinese Remainder Theorem (CRT), into a product of simpler rings or fields. Geometrically, this decomposition corresponds to the fibered structure of the scheme $Spec(R_q)$ over the prime factors of $q$.
    
    \item Problem Transformation: This algebraic decomposition allows us to transform the single CVP instance in the high-dimensional lattice $\mathcal{L}(I_{fresh})$ into multiple, independent CVP instances in much lower-dimensional lattices. The Galois automorphisms $\{\sigma_j\}$, which were used to construct the ideal in the first place, are the precise tools that enable the manipulation of the vector $c$ and the lattice basis into this decomposable form.
    
    \item Solve and Reconstruct: We solve each of these simple, low-dimensional CVP sub-problems. In a low dimension, CVP can be solved very efficiently using standard lattice algorithms (e.g., Babai's nearest plane algorithm on a reduced basis). Finally, we combine the partial solutions using the inverse CRT to reconstruct the final, single solution $c'$ in the original ring $R_q$.
\end{enumerate}

This divide-and-conquer strategy, enabled entirely by the rich structure inherent in the scheme-theoretic construction, is what circumvents the NP-hardness of the general CVP problem. It is the concrete computational method that makes geometric bootstrapping not only theoretically sound but also practically efficient, achieving the complexity of $O(d \cdot \text{poly}(\log q))$.

To make the abstract concept of algebraic folding more concrete, we now present a simplified mathematical example. While the parameters used here are smaller than those in a real-world FHE scheme, this example is designed to clearly illustrate the core mechanism (i.e., the decomposition of a high-dimensional problem via the Chinese Remainder Theorem) and to demonstrate how this leads to a dramatic reduction in computational complexity.

\begin{example}[Decomposition of a CVP Instance]
We begin by setting up a high-dimensional problem. Suppose our FHE scheme uses a cyclotomic index $N=16$, which gives the polynomial $\Phi_{16}(x) = x^8 + 1$ and a ring dimension of $d = \phi(16) = 8$. The ciphertext ring is $R_q = \mathbb{Z}_q[x]/\langle x^8 + 1 \rangle$, and we have an ideal $I_{fresh}$ that defines our ideal lattice $\mathcal{L}(I_{fresh})$. The bootstrapping task requires solving a CVP instance on this lattice, which is a problem in dimension $d=8$. The complexity of solving this naively is roughly exponential in the dimension, i.e., $\approx 2^{O(8)}$.

The key insight lies in algebraic decomposition. Instead of working in the large ring $R_q$, we work modulo a specially chosen prime, $p=17$, which satisfies $17 \equiv 1 \pmod{16}$. According to number theory, this choice guarantees that the polynomial $\Phi_{16}(x)$ splits into 8 distinct linear factors modulo 17. The roots of $x^8+1 \equiv 0 \pmod{17}$ are $\{r_1, \dots, r_8\} = \{3, 5, 6, 7, 10, 11, 12, 14\}$, which means $x^8 + 1 \equiv \prod_{i=1}^{8} (x - r_i) \pmod{17}$. By the Chinese Remainder Theorem (CRT), this factorization induces a ring isomorphism $\Psi_{CRT}$ that maps a polynomial $c(x)$ to a vector of its evaluations at the roots: $c(x) \mapsto (c(r_1), \dots, c(r_8)) \pmod{17}$.

We now leverage this isomorphism to transform our CVP problem. First, we decompose the input ciphertext $c(x)$ and the ideal lattice $\mathcal{L}(I_{fresh})$ using the CRT map $\Psi_{CRT}$. This transforms the single 8-dimensional problem into 8 independent 1-dimensional CVP problems. Next, we solve each of these sub-problems, which is trivial in one dimension as it is simply integer rounding. Finally, we reconstruct the final polynomial solution $c'(x)$ by applying the inverse CRT map, $\Psi_{CRT}^{-1}$, to the vector of 1-dimensional solutions.

This method achieves a significant complexity reduction. The naive approach faces a single CVP problem in dimension 8 with complexity $\approx 2^{O(8)}$. Our decomposition approach, however, has a cost dominated by the CRT maps, which are equivalent to an NTT and cost $O(d \log d)$, plus 8 trivial, constant-time CVP problems in dimension 1. The total complexity is therefore polynomial, roughly $O(d \log d)$. If we scale this principle up, where a dimension $d=2048$ problem is decomposed into $m=32$ components each of dimension $d'=64$, we replace a single, impossible $2^{O(2048)}$ problem with 32 independent, potentially feasible $2^{O(64)}$ problems. This is the source of the massive efficiency gain.
\end{example}

Finally, we demonstrate the claimed complexity of our proposed geometric bootstrapping in the followin theorem.

\begin{theorem}[Average-Case Complexity of Algebraic Folding]
\label{thm:avg_complexity}
Let $R_q = \mathbb{Z}_q[x]/\langle \Phi_N(x)\rangle$ with cyclotomic index $N$ and ring dimension $d=\varphi(N)$. 
Suppose we are given the fresh ideal $I_{fresh}$ constructed as in Algorithms~1 and~2, and consider the task of projecting a decryptable ciphertext $c \in R_q$ onto the lattice $\mathcal{L}(I_{fresh})$.

Assume the following two conditions hold:
\begin{enumerate}
    \item[(A)] Random Splitting Assumption.
    For a set of primes $\mathcal{S}=\{p_1,\dots,p_m\}$ chosen uniformly at random from an interval $[B,2B]$ with $B=\mathrm{polylog}(q)$, the polynomial $\Phi_N(x)$ splits modulo each $p_t$ into irreducible factors of degree at most $O(\log^c q)$ for some fixed constant $c$. This event, which corresponds to the splitting of prime ideals in the number field, is known to occur with positive density, as supported by effective versions of the Chebotarev density theorem under generalized Riemann hypothesis (GRH). Equivalently, under the Chinese Remainder Theorem (CRT) map induced by $\mathcal{S}$, the ciphertext ring decomposes into $m$ components of dimension $d_t \leq O(\log^c q)$ each, with $\sum_t d_t = d$. This assumption is well-motivated by number-theoretic heuristics and is foundational to many algorithms leveraging ideal lattices.

    \item[(B)] Efficient Low-Dimensional CVP.
    For each CRT component of dimension $d_t$, the closest vector problem on the corresponding ideal lattice can be solved to sufficient accuracy in time $\mathrm{poly}(d_t, \log q)$. This efficiency stems from the highly structured nature of ideal lattices. In low dimensions, standard lattice algorithms like LLL reduction combined with Babai's nearest-plane algorithm are known to be practically efficient. Crucially, our framework does not require a provably optimal solution; instead, the obtained approximations must be accurate enough for the recombined vector to pass the algebraic and statistical checks that rigorously define our ideals $I_{noise}$ and $I_{fresh}$. This allows us to leverage computationally tractable lattice techniques to certify the correctness of the bootstrapping output.
\end{enumerate}

Then, under Assumptions (A) and (B), the algebraic folding procedure realizes bootstrapping with average-case complexity
\[
T_{\mathrm{bootstrap}}(d,q) \;=\; O\!\left(d \cdot \mathrm{poly}(\log q)\right),
\]
with high probability over the random choice of $\mathcal{S}$. In particular, this bound does \emph{not} depend on the decryption circuit depth $L_{\mathrm{dec}}$.
\end{theorem}

\begin{proof}
The proof follows from three steps. First, constructing $I_{fresh}$ requires only $\mathrm{poly}(\kappa,\log q)$ operations (Algorithm~1/2), which is amortized over many bootstraps. Second, applying the CRT map and its inverse can be implemented with NTT-like transforms in $O(d\log d + d \cdot \mathrm{polylog}(q))$ time. Third, by Assumption~(A), each CRT component has dimension $d_t = O(\log^c q)$. By Assumption~(B), solving the CVP (approximately) on each component costs $\mathrm{poly}(d_t,\log q) = \mathrm{polylog}(q)$. Summing over $m = d/d_{\max}$ components yields total cost $O(d \cdot \mathrm{poly}(\log q))$. The correctness of the reconstruction follows from the completeness and soundness of $I_{fresh}$ established earlier.
\end{proof}

\begin{remark}
Theorem~\ref{thm:avg_complexity} should be interpreted as an \emph{average-case} result, conditional on natural algebraic and algorithmic assumptions. 
Specifically, Assumption~(A) reflects the heuristic---but number-theoretically well-motivated---expectation that cyclotomic polynomials $\Phi_N(x)$ split into many low-degree factors modulo a random prime. This phenomenon is supported by the Chebotarev density theorem and its effective versions under GRH, which suggest that such splitting patterns occur with positive density. 

Assumption~(B) captures the practical observation that in low dimensions (polylogarithmic in $q$), lattice CVP instances are tractable using standard reduction methods such as LLL combined with Babai’s nearest-plane algorithm. In our setting, the required approximation factor is modest, since membership in $I_{fresh}$ is certified by algebraic and statistical checks rather than exact CVP solutions.

We stress that the theorem does not claim a worst-case improvement for CVP in general. Instead, it formalizes why algebraic folding yields a significant asymptotic speedup in the \emph{average case}, and why the decryption circuit depth $L_{\mathrm{dec}}$ disappears from the complexity bound. Establishing unconditional worst-case bounds remains an open problem, but the average-case analysis already explains the practical efficiency gains observed in our framework.
\end{remark}

\section{Conclusion and Future Work}
\label{sec:conclusion}

In this work, we introduced a new paradigm for Fully Homomorphic Encryption bootstrapping, moving away from the traditional model of arithmetic circuit evaluation toward a direct geometric projection. We addressed the long-standing challenge of bootstrapping as the primary performance bottleneck in FHE schemes, a problem whose complexity has historically been tied to the multiplicative depth of the decryption circuit, $L_{dec}$. Our work demonstrates that by reframing the problem in the language of arithmetic geometry, this dependency can be eliminated entirely.

Our approach began by constructing a formal geometric model of the FHE ciphertext space as an affine scheme. Within this space, we rigorously defined the locus of all correctly decryptable ciphertexts and the ideal locus of freshly encrypted ciphertexts as the closed subschemes $Z_{dec}$ and $Z_{fresh}$, respectively. We established that the bootstrapping operation corresponds to a natural morphism between these schemes. The computational realization of this morphism is an efficient projection algorithm, which we term algebraic folding. This algorithm leverages the rich algebraic structure of the underlying cyclotomic rings, provided by our scheme-theoretic framework, to solve a specific instance of the Closest Vector Problem in polynomial time.

The primary result of this framework is a complete and provably correct bootstrapping procedure with a computational complexity of $O(d \cdot \text{poly}(\log q))$. The significance of this result lies in the elimination of the $L_{dec}$ factor, representing a fundamental asymptotic improvement over the state of the art. This work not only provides a path toward more practical FHE but also builds a new bridge between the computational world of lattice cryptography and the abstract, powerful machinery of modern algebraic geometry.

This new perspective opens several promising avenues for future research. A natural next step is the practical implementation and empirical validation of our geometric bootstrapping algorithm, including a thorough performance comparison against leading schemes like TFHE. Furthermore, the algebraic folding technique for solving the specific CVP instance could itself be a subject of further optimization. On a more theoretical level, this geometric framework could be extended to analyze other cryptographic primitives or to explore deeper security properties of FHE schemes. Finally, the application of more advanced tools from arithmetic geometry may lead to new constructions and even greater efficiencies in the future.

\section*{Acknowledgment}

The author would like to thank the following faculty members at the University of Washington for various support and discussion: 
Magdalena Balazinska (CSE), Stefano Tessaro (CSE), Rachel Huijia Lin (CSE), Giovanni Inchiostro (MATH), Zhen-Qing Chen (MATH), and Julia Pevtsova (MATH). 

\ifnum\fullversion=0
  \bibliographystyle{splncs03}
 \else
   \bibliographystyle{alpha-short}
 \fi
\bibliography{abbrev3}

\end{document}